%% file: main.tex
\documentclass[sigconf]{acmart}

\AtBeginDocument{%
  \providecommand\BibTeX{{%
    Bib\TeX}}}

\setcopyright{acmcopyright}
\copyrightyear{2024}
\acmYear{2024}
\acmDOI{XXXXXXX.XXXXXXX}

\acmConference[ICSE 2024]{46th International Conference on Software Engineering}{April 2024}{Lisbon, Portugal}
\acmPrice{15.00}
\acmISBN{978-1-4503-XXXX-X/18/06}

\usepackage{amsmath,amsfonts}
\usepackage{algorithmic}
\usepackage{graphicx}
\usepackage{textcomp}
\usepackage{xcolor}
\usepackage{colortbl}
\usepackage{fontawesome}
\usepackage{etex}
\usepackage{booktabs}
\usepackage[utf8]{inputenc}
\usepackage[ruled,vlined]{algorithm2e}
\usepackage[T1]{fontenc}
\usepackage{microtype}
\usepackage{graphicx}
\usepackage{paralist}
\usepackage{tabularx}
\usepackage{soul}
\usepackage{balance}
\usepackage{multicol}
\usepackage{multirow}
\usepackage{pbox}
\usepackage{enumitem}
\usepackage{lscape}
\usepackage{pifont}
\usepackage{xspace}
\usepackage{url}
\usepackage{tikz}
\usepackage{float}
\usepackage[TABBOTCAP]{subfigure}
\usepackage{ragged2e}
\usepackage{fontawesome}
\usepackage[figuresright]{rotating}
\usepackage{bbding} 

\usepackage{adjustbox}
\newcommand{\rev}[1]{\textcolor{black}{#1}}
\newcommand*\circled[1]{\tikz[baseline=(char.base)]{
		\node[shape=circle,fill,inner sep=0.8pt] (char) {\textcolor{white}{#1}};}}

\newcommand{\ie}{\emph{i.e.,}\xspace}
\newcommand{\eg}{\emph{e.g.,}\xspace}
\newcommand{\etc}{etc.\xspace}
\newcommand{\etal}{\emph{et~al.}\xspace}
\newcommand{\secref}[1]{Section~\ref{#1}\xspace}

\newcommand{\figref}[1]{Fig.~\ref{#1}\xspace}

\newcommand{\tabref}[1]{Table~\ref{#1}\xspace}

\newcommand{\metric}{\texttt{SIDE}\xspace} 

\newboolean{showcomments}

\definecolor{lightgrey}{rgb}{0.925, 0.925, 0.925}
\sethlcolor{lightgrey}

\setboolean{showcomments}{true}

\ifthenelse{\boolean{showcomments}}
  {\newcommand{\nb}[2]{
    \fbox{\bfseries\sffamily\scriptsize#1}
    {\sf\small$\blacktriangleright$\textit{#2}$\blacktriangleleft$}
   }
  }
  {\newcommand{\nb}[2]{}
  }

\def\BibTeX{{\rm B\kern-.05em{\sc i\kern-.025em b}\kern-.08em
    T\kern-.1667em\lower.7ex\hbox{E}\kern-.125emX}}
\begin{document}

\title{Evaluating Code Summarization Techniques:\\A New Metric and an Empirical Characterization}


\author{Antonio Mastropaolo}
\email{antonio.mastropaolo@usi.ch}
\affiliation{%
	\institution{SEART @ Software Institute, \\Universit\`a della Svizzera Italiana}
	\city{Lugano}
	\state{Switzerland}
	\country{CH}
}

\author{Matteo Ciniselli}
\email{matteo.ciniselli@usi.ch}
\affiliation{%
	\institution{SEART @ Software Institute, \\Universit\`a della Svizzera Italiana}
	\city{Lugano}
	\state{Switzerland}
	\country{CH}
}

\author{Massimiliano Di Penta}
\email{dipenta@unisannio.it}
\affiliation{%
	\institution{Dept. of Engineering, \\University of Sannio}
	\city{Benevento}
	\state{Italy}
	\country{IT}
}

\author{Gabriele Bavota}
\email{gabriele.bavota@usi.ch}
\affiliation{%
	\institution{SEART @ Software Institute, \\Universit\`a della Svizzera Italiana}
	\city{Lugano}
	\state{Switzerland}
	\country{CH}
}

\begin{abstract}
Several code summarization techniques have been proposed in the literature to automatically document a code snippet or a function. Ideally, software developers should be involved in assessing the quality of the generated summaries. However, in most cases, researchers rely on automatic evaluation metrics such as BLEU, ROUGE, and METEOR. These metrics are all based on the same assumption: The higher the textual similarity between the generated summary and a reference summary written by developers, the higher its quality. However, there are two reasons for which this assumption falls short: (i) reference summaries, \eg code comments collected by mining software repositories, may be of low quality or even outdated; (ii) generated summaries, while using a different wording than a reference one, could be semantically equivalent to it, thus still being suitable to document the code snippet. \rev{In this paper, we perform a thorough empirical investigation on the complementarity of different types of metrics in capturing the quality of a generated summary. Also, we propose to address the limitations of existing metrics  by considering a new dimension, capturing the extent to which the generated summary aligns with the semantics of the documented code snippet, independently from the reference summary. To this end, we present a new metric based on contrastive learning to capture said aspect. We empirically show that the inclusion of this novel dimension enables a more effective representation of developers' evaluations regarding the quality of automatically generated summaries.}
\end{abstract}



\maketitle

\input{introduction}
\input{related}
\input{approach}

\input{design}

\input{results}
\input{threats}
\input{conclusion}
\input{availability}

\section*{Acknowledgment}
This project has received funding from the European Research Council (ERC) under the European Union's Horizon 2020 research and innovation programme (grant agreement No. 851720).
Massimiliano Di Penta acknowledges the Italian PRIN 2020 Project EMELIOT ``Engineered MachinE Learning-intensive IoT system'',  ID 2020W3A5FY.
\newpage

\bibliographystyle{ACM-Reference-Format}
\bibliography{main}

\balance

\end{document}

%% file: introduction.tex
\section{Introduction} \label{sec:intro}

Program comprehension can take up to 58\% of developers' time \cite{Xia:tse2018}. Code comments are considered the most important form of documentation in this activity \cite{deSouza:2005}. Indeed, several studies provided evidence of the higher understandability of commented code as compared to non-commented code \cite{Woodfield1981,Tenny:tse1988}.

Despite the undisputed importance of code comments, developers do not always carefully comment code, or  update existing comments in response to code changes \cite{Steidl2013}. This may result in a lack of documentation \cite{deSouza:2005,spinellis:ie2010} and/or in outdated code comments \cite{Fluri:wcre07,Fluri:SQJ09,Linares:ASE15,Wen:icpc2019}. To support developers in such a task, researchers proposed code summarization techniques \cite{Rahman:SCAM15,Rodeghero:icse17,Haiduc:wcre2010,sridhara2011automatically,wong2015clocom, wong2013autocomment,moreno2013automatic,sridhara2011automatically,McBurney:tse2016,iyer:acl,allamanis2016convolutional,aghaj:2019a,LeClair:icse2019,Hu:emse2020,haque:2020,Zhang:icse2020}. These approaches take as input a code component to document (\eg a code function, or an entire class) and provide as output a natural language summary describing the code. The underlying technique can range from pre-defined templates properly filled via code analysis to the most recent techniques exploiting deep learning models trained on $\langle C,S \rangle$ pairs mined from software repositories, where $C$ represents the code to document and $S$ the original summary (comment) written by developers.

Empirically evaluating the quality of code summaries generated by these approaches is far from trivial. Indeed, assessing the extent to which a natural language text represents a good summary for a code component would require human (developers) judgment. Given the difficulties of running large-scale evaluations with developers, the software engineering community borrowed evaluation metrics from the Natural Language Processing (NLP) field. These include (but are not limited to) BLEU \cite{papineni2002bleu}, ROUGE \cite{lin2004rouge}, and METEOR \cite{meteor}. These metrics have been originally designed to act as a proxy for the quality of automatically generated text (\eg a translation) by comparing it with a reference (expected) text: The higher the words' overlap between the generated and the reference text, the higher the assessed quality. In NLP, the reference text is usually a collection of valid outcomes (\eg a set of valid translations for an input sentence) and the generated text is compared (in terms of word overlap) with all possible outcomes to obtain a better assessment. Also, the quality of the reference texts used for the evaluation is usually guaranteed by multiple human evaluators.

When adopting such metrics for code summarization, the generated summary is contrasted against a single reference text, usually being the original comment written by developers for the code provided as input, which can be easily mined from software repositories. However, two important issues emerge when using the NLP metrics in this context. First, there is no guarantee that the reference text is of high quality, as also demonstrated by empirical studies documenting quality issues in code comments (see \eg \cite{Fluri:wcre07,Linares:ASE15,Wen:icpc2019}). Thus, computing a word overlap between the generated and the reference summary may provide misleading indications of the actual quality of the generated summary. Second, the mined comment is only one of the possible ways to summarize the related code. 
Metrics based on word overlap penalize generated summaries for being different but semantically equivalent to the reference one, thus again not being good proxies for the summary quality.

The software engineering research community is well aware of the limitations of these metrics \cite{roy:fse2021,haque:icpc2022}. In response to the second limitation previously discussed (\ie capturing similarity between summaries using different wording but being semantically equivalent), Haque \etal \cite{haque:icpc2022} proposed the usage of word/sentence embeddings to properly capture the semantic similarity across summaries, moving from word overlapping to word-similarity measurement. While the authors show that these metrics better correlate with the human judgment of code summary quality as compared to word-overlapping metrics (\eg BLEU, ROUGE, METEOR), they still suffer of the first limitation we discussed, \ie \emph{a high similarity to a low-quality reference summary may provide a misleading good assessment of a generated summary}.

\rev{In this paper, we argue that an important factor in the assessment of summary quality is currently ignored by the state-of-the-art metrics: ``The suitability of the generated summary for the code to document, independently from the original comment written by developers''. To provide evidence of that, we present an empirical study that analyzes to what extent, different metrics to assess automatically-generated code summaries correlate, and complement with each other in explaining human assessment of such summaries. More importantly, we check for the complementarity of metrics capturing the new dimension with respect to others.
}
We rely on a dataset by Roy \etal \cite{roy:fse2021}, featuring more than 5k human evaluations of automatically generated summaries. We relate such evaluations to the quality assessment provided by different families of metrics. To capture the suitability of the generated summary for the code to document we experiment with (i) a simple approach relying on the word overlap between the summary and the code \cite{Steidl:icpc2013}; (ii) a deep learning-based approach exploiting embeddings obtained via a model pre-trained on code \cite{wang2023codet5};  and (iii) \metric (\textbf{S}ummary al\textbf{I}gnment to co\textbf{D}e s\textbf{E}mantics), a new metric leveraging contrastive learning \cite{schroff:cvpr2015} to model the characteristics of suitable and unsuitable code summaries for a given code. Our results show that (i) focusing on assessing the suitability of the generated summary for the documented code ignoring the reference summary allows to capture orthogonal aspects of summary quality as compared to state-of-the-art metrics, such as BLEU, ROUGE, METEOR, Jaccard similarity, as well as metrics based on word/sentence embeddings; and (ii) \metric is the metric having the strongest correlation with human judgment of code summary quality. We also show that \metric can be combined with state-of-the-art metrics to provide a more comprehensive assessment of code summary quality.

%% file: related.tex
\vspace{-0.1cm}
\section{Related Work} \label{sec:related}

This work relates to studies investigating factors impacting the soundness of evaluations performed to assess code summarization techniques (\secref{sub:sotaStudies}). Also, as \metric can be used
 to assess the quality of a code summary for a given code snippet, 
 we also discuss metrics capturing code comment quality (\secref{sub:sotaCommentQuality}).
\vspace{-0.2cm}
\subsection{Evaluating Code Summarization Techniques}
\label{sub:sotaStudies}

LeClair and McMillan \cite{leclair:2019} discuss the implications of different preprocessing choices made when preparing the datasets used for the evaluation of code summarization techniques, \eg splitting the training/test datasets by function \emph{vs} by project. 

They show that the impact of these choices can be drastic (up to $\pm$33\% of performance), and advocate for more standard dataset preparation procedures aimed at increasing results reproducibility.

Stapleton \etal \cite{Stapleton:icpc2020} present a study involving 45 university students and professional developers, in which participants were asked to perform program comprehension tasks on functions documented by human-written or automatically generated summaries. 
They found that participants performed significantly better with human-written summaries. Also, they show that the BLEU and ROUGE metrics computed on the automatically generated summaries do not correlate with the participants' performance in comprehending the documented code.
This poses questions on the suitability of these metrics to assess the quality of code summaries, at least when they are used to support program comprehension.

Gros \etal \cite{Gros:ase2020} discuss the basic assumption on which several code summarization techniques are built: The code summarization problem resembles the ``natural language translation'' problem, meaning that it can be seen as a ``code-to-natural language translation''. The validity of such an assumption would imply 
the possibility to reuse metrics successfully applied to assess the performance of automated natural language translations (\eg BLEU) in the context of code summarization. The empirical findings by Gros \etal show, however, that the two problems are substantially different. For example, while there is a strong input-output dependence in natural language translation (\ie similar input sentences result in similar output translations), this is not necessarily the case for code summarization (\ie similar functions could be commented in completely different ways). 

Roy \etal \cite{roy:fse2021} focus on the interpretation of metrics used in code summarization, and in particular on BLEU, ROUGE (in several different variants), METEOR, chrF \cite{popovic2015chrf}, and BERTScore \cite{zhang2019bertscore}. They conducted a study with researchers and practitioners in the software engineering community asking them to assess the quality of 36 summaries associated with 6 code snippets (a Java method). Each snippet had six summaries associated, one being the reference summary (\ie the one written by the original developers) and five resulting from different code summarization techniques \cite{leclair2019neural,haque:2020,alon2018code2seq,xu2018graph2seq,vaswani:nips2017}. The quality assessment has been performed using an ordinal scale in the range [0, 5]  \cite{Oppenheim:1992} (the higher the better) and focusing on three different aspects of each summary: conciseness, fluency, and content adequacy. Participants were not aware of which summaries were automatically generated and which, instead, represented the reference summary. Overall, they collected 226 surveys, for a total of 6,253 evaluations (not all participants fully completed the survey). By computing the above-listed evaluation metrics for the generated summaries and comparing them with the human assessment, Roy \etal found that metric improvements of less than two points do not indicate any meaningful difference in the quality of the summaries generated by two techniques. For higher differences metrics such as METEOR become reliable proxies for differences in summary quality, while others such as BLEU remain unreliable. We reuse the dataset by Roy \etal to assess the extent to which \metric and other metrics can be used to assess generated code summaries.

In a subsequent study, Hu \etal \cite{Hu:tosem2022} confirmed METEOR as the metric having the strongest correlation with the human judgment of code summary quality. Still, the achieved correlation is lower than that observed by human raters.
Haque \etal \cite{haque:icpc2022} criticize the use of word overlap metrics such as BLEU and ROUGE for assessing the quality of automatically generated summaries. They observe that using word overlap ignores the fact that (i) not all words have the same importance in a sentence; and (ii) the usage of synonyms in the generated summary (\eg using ``delete'' instead of ``cancel'' as in the reference summary) penalizes its quality assessment, which is conceptually wrong. Starting from this observation, they propose the usage of ``semantic similarity'' metrics based on word/sentence embeddings, which can better capture the similarity between the generated and the reference summaries. They then conducted a study involving 30 professional programmers, each of which was asked to evaluate the quality of a single summary associated with 210 Java methods. 
Their findings show that embedding-based metrics such as SentenceBERT \cite{reimers2019sentence} might be more suitable as quality assessment metrics for code summaries as compared to word overlapping metrics such as BLEU. In our study, we also consider all code summary quality metrics used in the work by Haque \etal
\vspace{-0.2cm}

\subsection{Assessing the Quality of Code Comments}
\label{sub:sotaCommentQuality}

Several works focused on the automated identification of inconsistent comments, namely comments that are not aligned with the code they document \cite{Tan:sosp2007,Tan:unenix2007,Tan:icst2012,Liu:compsac2018,Wang:ieee2019}. 
While these techniques are extremely valuable, they are not suited for the quality assessment of an automatically generated summary for the following reasons. First, their output is usually a binary classification indicating whether the comment is consistent or inconsistent, lacking one of the key characteristics of a metric, such as the possibility to compare and rank instances. Second, they focus on a specific aspect of low-quality comments, namely its inconsistency (\eg the comment documents a parameter no longer present in the method). However, a ``consistent comment'' might still be of low quality if it is too verbose, difficult to read, \etc As we will detail in \secref{sec:approach}, \metric (i) provides a continuous score ranging between [-1, 1], thus allowing comparing and ranking instances; and (ii) being trained via contrastive learning, learns the characteristics of suitable and unsuitable code summaries for a given code rather than relying on manually-crafted definitions of low-quality comments.

Khamis \etal \cite{Khamis:2010} propose JavadocMiner, a tool measuring several metrics which the authors consider important for the quality of Javadoc comments, \eg readability indexes, number of nouns in the comment, whether the Javadoc documents code entities (\eg parameters, return types) are not implemented in the documented code, \etc This approach exploits Javadoc-specific heuristics which cannot be generalized to the unstructured comments usually generated by code summarization techniques. Also, the quality-related metrics have been defined by the authors and never empirically evaluated against the developers' perception of comment quality. 

Steidl \etal \cite{Steidl:icpc2013} propose a machine learning-based approach to classify code comments into seven  categories (\ie copyright, header, member, inline, section, code, and task). Also, they propose a metric named c\_coeff to identify low-quality comments. The metric is based on the percentage of words in a comment that is similar to words in the code, where two words are considered similar if they have a Levenshtein distance lower than two (\ie at most one character must be changed to convert one word into the other). 
We consider c\_coeff in our study as a possible metric to capture the suitability of the generated summary for the documented code, showing that \metric better captures summary quality as perceived by developers.

\vspace{-0.2cm}

%% file: approach.tex
\section{SIDE} \label{sec:approach}
We present \metric, our novel metric to assess whether a natural language text represents a suitable summary for a given code. First, we provide background information about the DL model on top of which \metric is based (\secref{sub:mpnet}), and the contrastive learning procedure used to train it (\secref{sub:c-learning}). Then, we describe the dataset used for the model's training (\secref{sub:ft-dataset}). Finally, \secref{sub:training-evaluation} provides the details (\eg parameters) of the training procedure.

\subsection{MPNet in a Nutshell}
\label{sub:mpnet}
MPNet (Masked and Permuted Pre-training for Language Understanding) \cite{song:nips2020} is a Transformer \cite{vaswani:nips2017} pre-trained model built on top of BERT \cite{devlin:naacl-hlt2019}. Before discussing its architecture, let us briefly introduce the notion of pre-trained models. Pre-trained Transformers achieved state-of-the-art results in several Natural Language Processing (NLP) tasks \cite{devlin:naacl-hlt2019,yang:nips2019,liu:arxiv2019,raffel:jmlr2019,guo:iclr2021,yang:nips2019,song:nips2020,lewis:acl2020,uddin:naacl-hlt2021,zhang:iclm2020}. The pre-training, together with the self-attention mechanism featured in the Transformer architecture \cite{vaswani:nips2017}, played a major role in these achievements. The idea of pre-training is to provide the model with general knowledge about a language of interest before specializing it for a specific task. For example, let us assume we want to create an English-to-French translator. The model can be pre-trained on a large amount of unlabelled English and French data (\eg articles extracted from Wikipedia) using a self-supervised training objective, such as masked language modeling (MLM). MLM consists in randomly masking a percentage of the tokens in a given (English or French) sentence asking the model to guess them. For example, applying MLM to a sentence $\langle$$T_1$, $T_2$, $T_3$, $T_4$, $T_5$$\rangle$ we could obtain $\langle$$T_1$, $M$, $T_3$, $M$, $T_5$$\rangle$ (\ie $T_2$ and $T_4$ have been masked). The input for the model is the masked sentence, while the expected output are $T_2$ and $T_4$ as replacements for the two masked tokens. The idea is that thanks to MLM, the model starts acquiring knowledge about the languages' structure, preparing it to be specialized (fine-tuned) for the task of interest, namely language translation. 

The MLM pre-training objective has been adopted by several Transformer architectures, such as BERT \cite{devlin:naacl-hlt2019}. However, MLM suffers from a limitation: It ignores the dependencies among the masked tokens, possibly limiting the learning of complex semantic relationships. To overcome this limitation, Yang \etal \cite{yang:nips2019} proposed in XLNet the usage of permuted language modeling (PLM) during pre-training. 
\rev{This forces the model to learn long-range relations between tokens by guessing the correct positioning of the tokens in the whole sentence.
	While the technical details can be found in the paper introducing the technique \cite{song:nips2020}, the basic idea is that at pre-training time the model is provided with a permuted sentence featuring masked tokens and, on top of that, with original positioning information, \ie what was the original position of the tokens in the sentence before permutation. MPNet managed to achieve new state-of-the-art results in several works from the NLP community \cite{pan:aaai2022,martin:kbs22,gao:iccc2021,huang:emnlp2021}, including those relying on contrastive learning as training procedure \cite{pan2022improved,wei2022semi}, and it is the pre-trained model we specialize for the task of classifying a textual summary as suitable or not for a given code.
}

In terms of architecture, MPNet builds upon the $BERT_{base}$ model, which comprises 12 transformer layers with a hidden size of 768, 12 attention heads, and a total of 110M trainable parameters. MPNet was pre-trained using the same corpora exploited for the training of RoBERTa \cite{liu:arxiv2019}, which includes  datasets such as Wikipedia and BooksCorpus \cite{zhu:iccv2015}, OpenWebText \cite{gokaslan:openweb2019}, CC-News \cite{ccnews}, and Stories \cite{ccstories}, summing up for a total of 160GB of textual data.


\subsection{Contrastive Learning}
\label{sub:c-learning}

Contrastive learning \cite{schroff:cvpr2015} allows DL models to learn an embedding space where similar sample pairs (\ie pairs sharing specific features) are clustered together while dissimilar pairs are set apart. In our context, we use contrastive learning to discriminate suitable \emph{vs} unsuitable summaries for a given source code snippet. To this aim, we need to show to the model both \emph{positive samples} (source code associated with suitable summaries) and \emph{negative samples} (source code associated with unsuitable summaries). 

Several contrastive representation learning losses have been proposed in the literature \cite{hadsell:cvpr2006,schroff:cvpr2015,huber:breakthroughs1992,kullback:courier1997}.  We employ the \emph{triplet loss} \cite{schroff:cvpr2015}, which has been shown to better encode the positive/negative samples as compared to other contrastive losses \cite{chopra2005learning}. The triplet loss function has been proposed by Schroff \etal \cite{schroff:cvpr2015} and introduces the concept of ``anchor''. Given an anchor $x$, a positive ($x^{+}$) and a negative ($x^{-}$) sample is selected, with the triplet loss which during training minimizes the distance between the $x$ and $x^{+}$, while maximizing the distance between $x$ and $x^{-}$.
In our case, the anchor is the code to document, with a suitable summary representing $x^{+}$ and an unsuitable summary representing $x^{-}$. In the following, we introduce the dataset used to fine-tune MPNet for the task of interest, explaining how we generate positive and negative samples.

\subsection{Fine-tuning Dataset}
\label{sub:ft-dataset}
We need to collect code instances paired with ``suitable'' and ``unsuitable'' code summaries. In our evaluation of \metric (\secref{sec:design}) we will exploit a dataset from the literature featuring developers' evaluation of code summaries for Java methods \cite{roy:fse2021}. 

Thus, we started by collecting Java methods accompanied by a textual summary. We exploited the CodeSearchNet dataset \cite{husain2019codesearchnet}, featuring $\sim$6M functions from various programming languages, including Java, and a subset of Java methods is accompanied by a Javadoc description. Lu \etal \cite{lu2021codexglue} observed that the original CodeSearchNet dataset featured instances potentially being problematic. Therefore, they created a curated version of the dataset excluding methods that cannot be parsed, and those paired with a Javadoc (usually, the method's summary) having its first sentence shorter than 3 or longer than 256 tokens, and featuring special tokens (such as $\langle img\rangle$), or not being written in English. Their refined Java subsection of the CodeSearchNet dataset comprises 181,061 pairs of $\langle$$method$, $summary$$\rangle$ which have been already split into three subsets: 164,923 training, 5,183 validation, and 10,955 testing. The $summary$ here is the first sentence of the Javadoc \rev{extracted as the first paragraph of the documentation (\ie the one delimited with the first period) \cite{lu2021codexglue}.} 
\rev{To increase the confidence in the quality of the exploited dataset and verify whether the sentences automatically extracted from the documentation actually represent summaries of the method, two of the authors independently inspected 100 randomly selected samples from the dataset, classifying their associated documentation (\ie the first automatically extracted first paragraph) as ``code summary'' or ``other''. The guideline was to classify it as a code summary if it summarizes the intent of the method (\eg comments documenting a self-admitted technical debt \cite{cunningham1992wycash} are tagged as ``other''). After solving 2 cases of disagreement, 95 of the inspected documentations were classified as actual summaries.}

The starting assumption is that the original summary written by the developers is a positive sample when paired with its associated method. This makes for 164,923 positive samples in our dataset. The same number of negative samples can be easily created by associating each method with a randomly selected summary from the training set (different from the original summary). The result is a dataset of 164,923 $\langle$$method$, $positive$, $negavite$$\rangle$ triplets featuring, for each method, a positive and a negative summary. The approach used to create the negative samples, while simple, may associate unrelated summaries to methods, simplifying the learning of the model, \ie it becomes rather easy to discriminate between positive and negative samples at training time. However, when assessing the quality of an automatically generated summary (our final goal for \metric), it is unlikely that the latter is completely unrelated to the input method, even when it is of low quality. Thus, we must train \metric so that it is able to identify as ``unsuitable'' for a given method even summaries which are plausible yet still suboptimal. These are known in the literature as hard-negative samples \cite{oh2016deep,schroff:cvpr2015}. 

To automatically generate these summaries, we conjecture that inner comments only documenting a subset of the method's statements are unsuitable as ``method's summary''. Nevertheless, they are still likely to be related to the method, certainly more than randomly selected summaries. Thus, we parsed the methods in the training set to extract all their inner comments and associated each inner comment to the set of statements it documents. For such association, we use the heuristics previously proposed in the literature, linking each inner comment to all following statements until an empty line or a closing curly bracket is reached \cite{chen:jss2019,huang:ist2020}. Comments reporting self-admitted technical debt \cite{Potdar:icsme2014} have been identified using keywords matching and excluded (we removed all comments including one of the following words: \emph{to-do, fix-me, todo, fixme, xxx, hackme, hack-me}). Indeed, these comments do not describe the code.  We then computed the percentage of statements in the method that each comment documents, only considering actual code statements (\ie excluding comments and blank lines). We consider a comment as a good hard-negative sample (\ie a plausible method description being, however, unsuitable as a method's summary) if it documents less than 25\% of the method's statements.  \rev{The choice of the 25\% threshold is motivated by the following assumption: if a comment describes at most one fourth of the statements in a method, it is unlikely that it can represent a comprehensive summary of it while still possibly documenting some of the responsibilities implemented in the method.}
We managed to create a hard-negative for $\sim$15\% (24,951) of the instances in our training dataset, since the others did not have any inner comment matching our selection procedure.

Our training dataset features 164,923 triplets generated using random negatives, and 61,001 triplets featuring instead hard-negatives, since each of the 24,951 methods for which we managed to create a hard-negative can contribute with more than a single instance.

\subsection{Training and Model Evaluation}
\label{sub:training-evaluation}


	MPNet has been trained for 10 epochs \rev{(which accounted for 141,205 training steps)} using a batch size of 16 and a maximum sequence length of 512 tokens. The sequence length acts on the input of the model (in our case, the concatenation of a method and its summary), cutting out longer sequences. 
	 \rev{This impacted only 4.27\% of the instances in our fine-tuning dataset.}
\rev{The learning rate has been warmed-up by taking into account the overall size of the training dataset, batch size and number of epochs.
	This strategy, which increases the learning rate from 0 to 2e-5 (default values when using the AdamW optimizer \cite{loshchilov:iclr19}) has been previously adopted when training BERT-based models using Sentence Transformers \cite{reimers2019sentence}
} 
The best-performing checkpoint \rev{which we found to be the last one saved (\ie after 141,205 training steps)} has then been selected as the one maximizing the following score (also used in previous work using contrastive learning \cite{zhang:iclr2020,saadany:arxiv2021}):

To reduce the chance of overfitting, we save checkpoints every 5k training steps while using a patience of 5. Specifically, we evaluate the model on the validation set (\secref{sub:ft-dataset}), which includes 5,183 positive samples (the original descriptions associated with each method) and the same number of negative samples randomly generated as explained in \secref{sub:ft-dataset}.
The best-performing checkpoint has then been selected as the one maximizing the following score (also used in previous work using contrastive learning \cite{zhang:iclr2020,saadany:arxiv2021}):
\vspace{-0.03cm}
\[
\frac{\sum_{i=1}^{N}   \text{CSpositive}_i - \text{CSnegative}_i}{N}
\]
\vspace{-0.03cm}
\noindent where $N$ represents the number of positive and negative samples (which in the evaluation set is guaranteed by construction to be the same), $\text{CSpositive}_i$ represents the cosine similarity returned by the model between the method $m_i$ and its ``positive'' summary, while $\text{CSnegative}_i$ represents the cosine similarity returned between method $m_i$ and its ``negative'' summary. A perfect model would return 1.0 for such a metric, reporting 1.0 as the similarity of all positive summaries and -1.0 as the similarity of all negative summaries.
We acknowledge that our evaluation set features randomly generated negative instances that, as such, are simple to identify by the model. However, such a procedure is just used to select the best-performing checkpoint for \metric, and not for its  empirical evaluation (described in \secref{sec:design}).

%
%



%% file: design.tex
\section{Study Design} \label{sec:design}

The \emph{goal} of this study is to evaluate metrics used in the literature to assess the quality of code summarization approaches, including our newly proposed metric \metric using contrastive learning. The \emph{quality focus} is the complementarity of the considered metrics with respect to others, and the extent to which the considered metrics would explain summary quality assessments performed by developers. The \emph{perspective} is of researchers wanting to define a framework for the assessment of code summary quality, that can be used, for example, to evaluate code (re)documentation approaches. 

The \emph{context} consists of (i) the dataset by Roy \etal \cite{roy:fse2021} featuring more than 5k developers' quality assessments of automatically generated summaries; and (ii) 40 summary evaluation metrics, including \metric. The study addresses the following research questions:\smallskip

\newcommand{\rqone}{To what extent different metrics to evaluate the quality of source code summarization correlate with each other?}
\newcommand{\rqtwo}{To what extent different metrics to evaluate the quality of source code summarization contribute to explain user-based evaluations?}

\textbf{RQ$_1$}: \emph{\rqone} \rev{Before analyzing how metrics contribute to explaining the quality of generated summaries, we analyze the extent to which they are related with each other, or, instead, capture different dimensions of the dataset variability.}
\smallskip
	
\textbf{RQ$_2$}: \emph{\rqtwo} \rev{After having identified a subset of unrelated metrics capturing the dataset variability, we analyze the extent to which  these metrics contribute to explaining different summarization quality indicators. This would aim at providing a quantitative indication not only of the complementarity of the different metrics but also of the importance of each metric to describe an extrinsic quality indicator.}

\subsection{Evaluation Dataset}
\label{sub:eval-dataset}
We use the dataset by Roy \etal \cite{roy:fse2021} described in \secref{sec:related}. As previously explained, this dataset features humans' evaluations of both automatically generated summaries for Java methods as well as of the original summaries written by the methods' developers. The latter are not suitable for our study. Indeed, to compute some of the evaluation metrics (\eg BLEU score), we need an automatically-generated summary to contrast against a reference summary. Thus, we remove from the set of 6,253 evaluations the 1,052 evaluations referring to the (manually written) original summaries, leaving us with the dataset of 5,201 evaluations. 

\subsection{Variable Selection}
\label{sub:variable-selection}

The 5,201 evaluations performed by developers \cite{roy:fse2021} concern three quality aspects of summaries, all rated on an ordinal scale from 0 to 5 \cite{Oppenheim:1992} (the higher the better):

\begin{itemize}
	\item \textbf{Conciseness:} Assesses the degree to which the summary contains unnecessary information.
	\item \textbf{Fluency:} Evaluates the continuity or smoothness rate in the generated summary.
	\item \textbf{Content Adequacy:} Assesses the extent to which the summary lacks information needed to understand the code.
\end{itemize}

In addition Roy \etal \cite{roy:fse2021} also collected a Direct Assessment (DA) score \cite{graham2013continuous}, expressed on a scale from 0 to 100, and providing an overall quality assessment of the summary. The three quality aspects (conciseness, fluency, and content adequacy), and the DA score represent the \emph{dependent variables} of our study.

Regarding the \emph{independent variables}---\ie automated evaluation metrics for code summaries---we considered, besides \metric, (i) all metrics in the work by Roy \etal \cite{roy:fse2021}, (ii) additional metrics used in the work by Haque \etal \cite{haque:icpc2022} (see \secref{sec:related}), (iii) the \texttt{c\_coeff} metric proposed by Steidl \etal \cite{Steidl:icpc2013}, and (iv) a baseline we define based on CodeT5+ \cite{wang2023codet5} (a Transformer pre-trained on code and English text) to compute the textual similarity between the generated summary and the code. The latter has the purpose to let us check the actual benefits (if any) brought by the contrastive learning we adopt in \metric as compared to other DL-based approaches.
The considered metrics, besides being a comprehensive set of those used in the literature to assess code summarization techniques, approach the quality assessment of code summaries in different ways. 

In particular, the ones inherited from the works by Roy \etal \cite{roy:fse2021} and Haque \etal \cite{haque:icpc2022} are conventional metrics, in the sense that they look at the similarity between the generated summary and the reference summary. 
These metrics range from very simple (\eg word overlapping between the two summaries), to more complex ones exploiting DL-based embeddings. Differently, c\_coeff \cite{Steidl:icpc2013}, CodeT5+, and \metric look for the similarity between the generated summary and the documented code. This is the first time such a dimension is considered in the assessment of code summarization techniques. While several approaches can be used to measure such a similarity, we opted for three metrics including a ``trivial'' solution (\ie the c\_coeff metric exploiting word overlap), a state-of-the-art Transformer pre-trained model (CodeT5+), and a contrastive learning-based solution being \metric. 
\rev{
	For what concerns \metric, we consider two of its variants: One exploiting hard-negatives in the training set (as extracted using the procedure described in Section \secref{sub:ft-dataset}) and one only including random negatives. This is basically an ablation study investigating the role played by the hard-negatives on the ability of \metric to assess the suitability of a summary for a given code.
}

 
\subsubsection{Words/characters-overlap based Metrics}
\hfill

\textbf{BLEU} (BilinguaL Evaluation Understudy) \cite{papineni2002bleu} measures the similarity between the candidate (predicted) and reference (oracle) summaries. Such a similarity assesses the overlap in terms between the two summaries and it is defined on a scale between 0 (completely different summaries) to 1 (identical summaries). We compute the BLEU score at the sentence level for various values of $n$, including $n$=\{1, 2, 3, 4\}. We also compute the BLEU-A being the geometric mean of the four BLEU variants we consider.

\textbf{METEOR} (Metric for Evaluation of Translation with Explicit ORdering)~\cite{meteor} is computed as the harmonic mean of unigram precision and recall, with recall being assigned a higher weight than precision. In contrast to BLEU, METEOR incorporates stemming and synonyms matching to align more closely with human perception of similarity between sentences. The METEOR score ranges from 0 to 1, with a score of 1 indicating two identical sentences.

\textbf{ROUGE} (Recall-Oriented Understudy for Gisting Evaluation)~\cite{lin2004rouge} is a set of metrics for evaluating both automatic summarization of texts and machine translation techniques. ROUGE metrics compare an automatically generated summary or translation with a set of reference summaries (typically, human-produced). Similarly to Roy \etal \cite{roy:fse2021}, we compute ROUGE-N(1-4), ROUGE-L, and ROUGE-W. ROUGE-N measures the number of matching $n$-grams between the generated summary and the reference summary with results reported in terms of recall, precision and f1-score.

\textbf{Jaccard}~\cite{jaccard} measures the degree of overlap between two sets of tokens (summaries in our case). It is calculated by dividing the size of their intersection by the size of their union, thus obtaining a value between 0 and 1 (the higher the Jaccard, the more similar the two summaries).

\textbf{chrF} (character $n$-gram F-score)~\cite{popovic2015chrf} measures the similarity between the generated and the reference summaries at the character level (rather than at the word-level as done by the above metrics), reporting the computed value using the F-score.

\textbf{c\_coeff} \cite{Steidl:icpc2013}, while still based on word overlap, focuses on the similarity between a summary and its associated code: It computes the percentage of words in a summary that is similar to words in the code, where two words are considered similar if they have a Levenshtein distance lower than two.

\subsubsection{Embedding-based Metrics}
\hfill

\textbf{TF-IDF} (Term frequency-inverse document frequency) \cite{ramos2003using} is a widely used term weighting schema assessing the importance of a word within a document collection. 
In our context, TF-IDF is used to compute the cosine similarity (TF-IDF\_CS) and the Euclidean distance (TF-IDF\_ED) between the terms' vectors representing the generated and the reference summary.

\textbf{BERTScore} \cite{zhang2019bertscore} computes sentence similarity using the embedding of the BERT model \cite{devlin:naacl-hlt2019}, which has been trained on English textual data. 
We report all the BERTScores which include, precision (BERTScore-P), recall (BERTScore-R), and F1-score (BERTScore-F1).

\textbf{SentenceBERT} \cite{reimers2019sentence} employs a siamese network architecture to generate fixed-length representations of sentences using BERT \cite{devlin:naacl-hlt2019} as a backbone to produce the encoding. The representations of the generated and the reference summary are compared via cosine similarity (SentenceBERT\_CS) and the Euclidean distance (SentenceBERT\_ED).

\textbf{InferSent} \cite{conneau2017supervised} relies on GloVe vectors \cite{pennington2014glove} as pre-trained word embeddings for the sentence pair. The embeddings are then passed through RNN encoder layers to obtain fixed-length vector representations for each sentence. 
Also in this case both the cosine similarity (InferSent\_CS) and the Euclidean distance (InferSent\_ED) are considered to contrast the generated and the reference summary.
  
\textbf{Universal Sentence Encoder (USE)} \cite{cer2018universal} employs transformer encoders to generate context-aware representations of words within a sentence by levering the self-attention mechanism.
Both USE\_CS (cosine similarity-based) and USE\_ED (Euclidean distance-based) are considered. 

\textbf{CodeT5+\_CS} \cite{wang2023codet5} exploits CodeT5+, a model pre-trained on code and natural language. We use the CodeT5+$_{base}$ variant ($\sim$220M trainable parameters) to compute the cosine similarity between the embeddings of the generated summary and the code to document. CodeT5+\_CS acts as a further baseline for \metric which has been fully trained via contrastive learning.

\medskip
  
%
%
%

\subsection{Analysis Methodology}
\label{sub:analysis}
In the following, we describe the study analysis methodology. The whole analysis has been performed using the \emph{R} \cite{R} statistical environment. For statistical tests, we assume a significance level $\alpha=0.05$.

To address RQ$_1$, we first analyze the correlation between different summary evaluation metrics. To avoid being constrained with linear relationships only, we leverage the non-parametric, Spearman's rank correlation \cite{Conover:1998}.  To show the correlation, we create a visual overview of correlation among metrics using the \emph{varclus} function part of the \emph{R} \emph{Hmisc} package \cite{hmisc}. The output of this procedure is a hierarchical clustering of variables (\ie our metrics), producing a dendogram (\ie the clustering tree) visualizing correlated metrics. 
\rev{Note that we do not use Spearman's correlation to select uncorrelated variables but, as explained below, the \emph{redun} procedure which allow accounting for multicollinearity other than collinearity.}

We complement the correlation analysis with a Principal Component Analysis (PCA) \cite{Jolliffe86}. PCA leverages Singular Value Decomposition (SVD) to describe the underlying data variance and covariance expressed as linear combinations of the considered variables. 

To avoid having the results of the PCA being affected by collinearity, we run a variable selection procedure using the \emph{redun} function from the \emph{Hmisc} package \cite{hmisc}. This function performs a stepwise removal of the independent variables determining how well each variable can be predicted by the remaining ones. The process starts by first removing the most predictable variable and continuing subsequently until no variable among the predictors can be predicted with a given (adjusted) $R^2$, which we set equal to 0.8. Then, before applying PCA, we re-scale the variables in the range [0,1] using a min-max re-scaling.
 
The PCA on the selected variables returns the eigenvectors related to each independent component, where the (absolute) values in correspondence of each metric indicate the importance of the metric for that component. Moreover, it returns the standard deviation captured by each principal component, which indicates the importance of the component itself. The PCA has been performed using the \emph{prcomp} function of the \emph{R} (default) \emph{stats} package.

\rev{To address RQ$_2$, we determine the extent to which different summary quality metrics can complement each other to explain extrinsic quality indicators provided by users. Given the (ordinal) scale of the dependent variables, and given that we cannot assume a linear relation between independent and dependent variables, this analysis has been performed by employing a multivariate logistic regression. A conventional logistic regression model is not suitable for our analysis, because the dependent variables are not dichotomous but, rather, expressed on a 0-5 ordinal scale \cite{Oppenheim:1992}, or on a 1-100 scale (in the case of DA score).} Therefore, we use an ordered  logistic regression. Given the $l$ levels of an ordinal variable, the ordinal logistic regression models $\ell$ different logits as:

\begin{equation}
ln \left( \frac{P(Y \le \ell)}{P(Y > \ell)} \right) = \xi_\ell - \eta_1 X_1 - \dots - \eta_n X_n\\
\end{equation}
\vspace{0.2cm}

\noindent where $Y$ is the model's dependent variable, $X_i$ the independent variables, $\eta_i$ their coefficients (estimates), and $\xi_\ell$ the intercept.

We use the \emph{polr} function from the \emph{MASS} package \cite{MASS}. The interpretation of the model estimates is similar to the ones of logistic regression, with the difference that, given an estimate $\eta_i$, the OR=$e^{\eta_i}$ indicates what are the increased odds of a unitary value increment for the dependent variable given a unitary increase of an independent variable. The application of the ordered logistic regression follows three subsequent steps, \ie (i) variable selection, (ii) variable re-scaling, and (iii) model building. 

For the variable selection, we use the output of the \emph{redun} procedure used in RQ$_1$. Then, given the selected variables, we re-scale them in the same range of the dependent variable being modeled. This makes the interpretation and comparison of the model's ORs easier. As for the dependent variables, we leave them as they are (\ie on a scale from 0 to 5 for \emph{conciseness}, \emph{fluency}, and \emph{content adequacy}, and on a scale from 1-100 for \emph{DA score}).

Finally, we apply the  \emph{polr} procedure, relating each dependent variable to all independent variables that were left after the \emph{redun} feature selection. We report (i) fitting diagnostics (in particular the Akaike Information Criterion - AIC value), (ii) independent variables estimates, OR, and significance \emph{p}-value adjusted, due to multiple factors, with the Benjamini-Hochberg correction \cite{bh}.\\

%% file: results.tex

\section{Results Discussion} \label{sec:results}


\subsubsection*{RQ$_{1}$: Correlation of different metrics to evaluate the quality of source code summarization}

\begin{figure*}
	\centering
		\caption{RQ$_1$ - Spearman's correlation clustering on the independent variables with {varclus}}
	\includegraphics[width=0.99\textwidth]{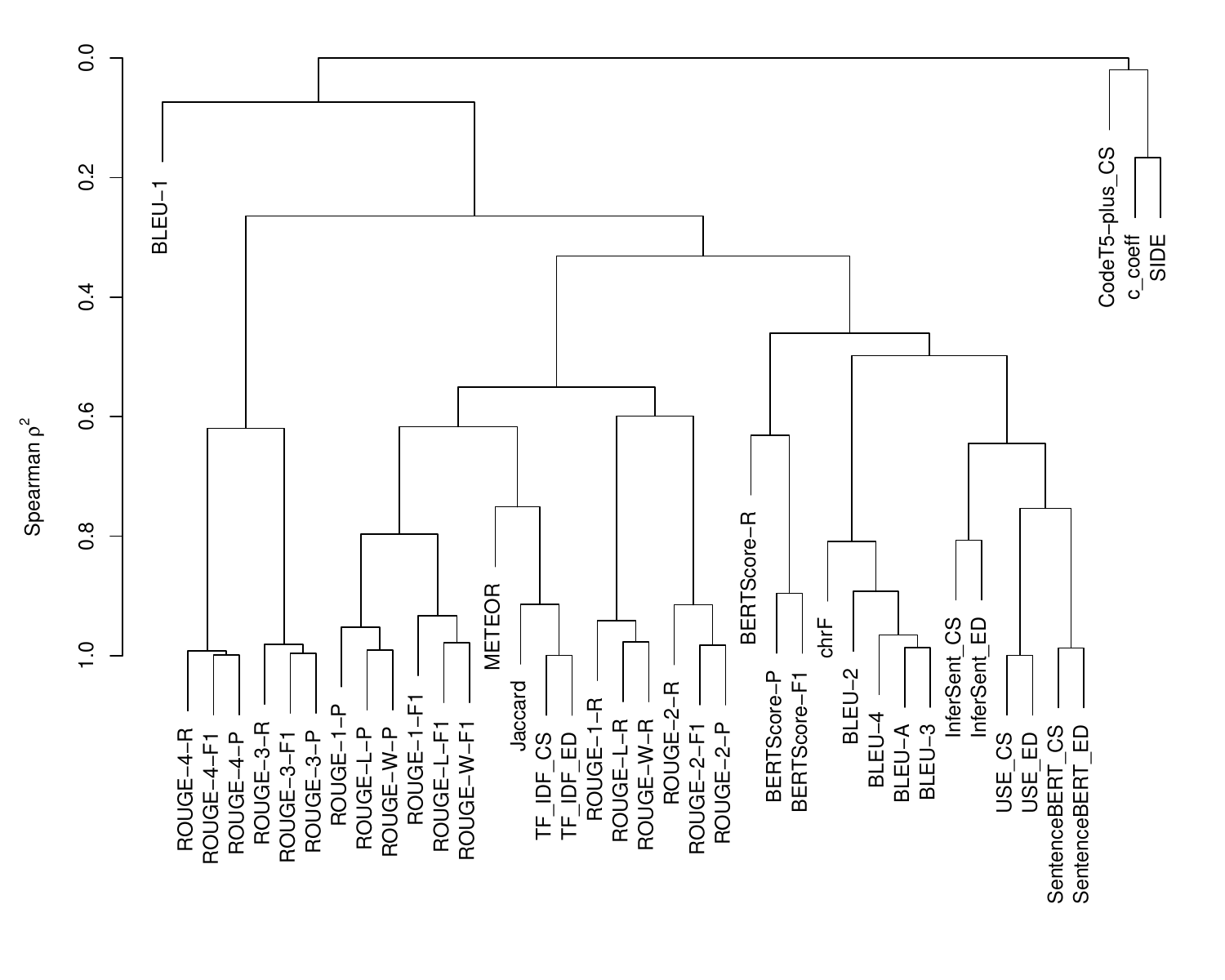}
	\vspace{-1cm}
	\vspace{-0.5cm}
	\label{fig:dendrogram}
\end{figure*}

\figref{fig:dendrogram} depicts the hierarchical clustering of the independent variables output of the \emph{varclus} procedure. Note that the values on the $x$-axis show the square of the Spearman's correlation (\eg a value of 0.64 indicates a correlation of $0.8^2=0.64$). 

The top-level fork of the tree splits the metrics into two different families. The first groups all conventional metrics capturing the similarity between a generated and a reference summary, while the second includes the three metrics focusing on the relevance of the generated comment for the documented code (\ie \texttt{CodeT5-plus\_CS}, \texttt{c\_coeff}, and \metric). This is a first indication of the fact that the two families of metrics focus on different aspects of summary quality.

Looking inside the sub-tree composed of conventional metrics, \texttt{BLEU-1} is not correlated with all others, not even with the other BLEU variants. While this may look counter-intuitive, \texttt{BLEU-1} focuses on single-word overlap, while its other variants look at $n$-grams overlap. This usually results in much higher values for the \texttt{BLEU-1} as compared to \texttt{BLEU-2}, \texttt{BLEU-3}, and \texttt{BLEU-4}.

The deeper we go into the tree, the more cohesive the clusters of metrics that can be observed. The three metrics looking at the documented code (\ie \texttt{CodeT5-plus\_CS}, \texttt{c\_coeff}, and \metric) are very high in the dendogram indicating that, while they stay apart from the other conventional metrics, the correlation among them is small, likely due to the different underlying solutions used to assess the relevance of a generated summary for a given code.

\input{tables/pca}

The next step in our data analysis is to use the PCA to identify the orthogonal components in the performed measurements. To run it, we first selected the metrics to consider using the \emph{redun} function as explained in \secref{sub:analysis}. This process resulted only in the 10 metrics listed in \tabref{tab:pca}, already providing a first indication of the fact that the other metrics considered in our study are redundant and highly correlated with at least one of the 10 selected. The latter include (i) four words/characters-overlap based metrics capturing the similarity between the generated and the reference summary (\ie \texttt{BLEU-1}, \texttt{ROUGE-1-P}, \texttt{ROUGE-4-R}, and \texttt{ROUGE-W-R}); (ii) three embedding-based metrics also focusing on the similarity between the generated and the reference summary (\ie \texttt{BERTScore-R}, \texttt{SentenceBERT\_CS}, and \texttt{InferSent\_CS}); and (iii) all three metrics focusing on the similarity/relevance of the generated summary to the documented code (\ie \texttt{c\_coeff}, \texttt{CodeT5-plus\_CS}, and \metric).

\tabref{tab:pca} reports the results of the PCA, with 10 principal components (PCs) identified. The first row indicates the proportion of variance captured by each PC. The higher such a value, the higher the variance in the dataset described by the PC. The second row reports the cumulative proportion of variance when considering the first $n$ PCs. For example, by just using five PCs, it is possible to capture 91\% of the variance in the data, \ie 55\% (PC1) + 16\% (PC2) + 8\% (PC3) + 7\% (PC4) + 4\% (PC5). The remaining 10 rows in \tabref{tab:pca} show the importance of each metric for each PC: The higher the absolute value, the higher the metric's contribution to that PC. For each PC, we highlight with a dark background the metric(s) contributing the most to it. In particular, we highlight the metric with the highest absolute value in the eigenvector and all those close to it (at most -5\% within the absolute value). The ``5\%'' choice is arbitrary and just meant to simplify the results' visualization and discussion highlighting all metrics similarly contributing to a PC.

PC1 is captured by conventional metrics looking for textual similarity between the generated and the reference summary (mostly \texttt{BERTScore} and \texttt{ROUGE}). PC2 is instead exclusively captured by \metric, suggesting the importance of including metrics considering the code to document when measuring summary quality. 

The second-highest value for PC2 is obtained by using \texttt{c\_coeff}, but it is substantially smaller than the one associated with \metric (0.47 \emph{vs} 0.79). For PC3, \texttt{c\_coeff} exhibits instead the highest eigenvector. 

The analysis conducted so far indicates that by just considering three metrics (\eg \texttt{BERTScore-R}, \metric, and \texttt{c\_coeff}) it is possible to capture 80\% of the variance in the data. 

Conventional metrics are also associated with PC4 (\texttt{ROUGE-4-R}), with \texttt{c\_coeff} being instead the one capturing PC5. Basically, among the top-5 PCs, 3 are mostly captured by metrics considering the code to document in the equation.

In summary, RQ$_1$ findings show that conventional metrics looking at the similarity between the generated and the reference summary do not correlate with those looking at the relevance/similarity of the generated summary for/with the documented code. Also, they only capture part of the variance in the data. 

This only indicates that the metrics considering the documented code (\eg \metric, or \texttt{c\_coeff}) capture orthogonal information as compared to the conventional ones, not whether or not they better capture code summary quality as perceived by developers. 

\subsubsection*{RQ$_{2}$: Contribution of different code summarization metrics in explaining user-based evaluations}


\textcolor{black}{Before discussing how various metrics to assess source code summarization impact user-based evaluations, we must anticipate that, in this section will focus exclusively on the outcomes achieved when training \metric using hard-negatives. Additional information about the conducted ablation study is provided in \secref{subsec:ablation}.
}

\tabref{tab:logistic} reports the results of the ordered logistic regression for each code summary quality attribute (dependent variables) manually evaluated by developers (\ie \emph{DA score}, \emph{content adequacy}, \emph{conciseness}, and \emph{fluency}).  We focus our discussion on the Odds Ratios (ORs) of the statistically significant $p$-values ($<$0.05), which are the ones highlighted with a black background. The interpretation of the ORs is as follows: they indicate the odds of a unitary increment in the dependent variable given a unitary increment in the independent variable. For example, if we look at the OR of \metric when considering the \emph{content adequacy} as a dependent variable (1.6265), it indicates $\sim$62\% higher odds of observing a unitary increment in the \emph{content adequacy} as perceived by developers for each unitary increment of the \metric value. Remember that both independent and dependent variables have been normalized on the same scale: In the case of \emph{DA score}, all variables are in the 1-100 range, while the three other variables are in the range 0-5, since we followed the scales used in the developers' evaluation. This also explains the lower ORs in the \emph{DA score} table. Indeed, a unitary increase on a 1-100 range has a different magnitude as compared to a unitary increase on a 0-5 scale.

\input{tables/logistic-model}

A first observation that can be made is that \metric is the metric having the highest OR independently from the considered dependent variable. Also, only five metrics obtained a significant $p$-value for at least one dependent variable (\metric is always among them). 

The ``overall'' quality of the summary (\ie \emph{DA score}) and its \emph{content adequacy} as perceived by developers is captured by five metrics: \metric, \texttt{c\_coeff}, \texttt{CodeT5-plus\_CS}, \texttt{SentenceBERT\_CS}, and \texttt{ROUGE-1-P}, with \metric and \texttt{c\_coeff} having the highest and second-highest OR, respectively, for both dependent variables. Still, conventional metrics looking at the similarity between a generated and a reference summary also play a role in capturing code summary quality as assessed through \emph{DA score} and \emph{content adequacy}.

When moving to the summary \emph{conciseness} and \emph{fluency}, \metric and \texttt{c\_coeff} confirm their strong relationship with the human assessment (with \metric being the best in the class), while the conventional metrics struggle in capturing these quality aspects, with the only exception being the \texttt{ROUGE-W-R} when looking at the \emph{conciseness}. 

Still, the OR for \metric is 1.3844 (\ie 38\% higher odds of a unitary increment in the human assessment of summary conciseness for a one unit increase of \metric), while for \texttt{ROUGE-W-R} is substantially lower (1.0954). The reason may lie behind the different assumptions made by the two families of metrics. To explain this point, let us take the example of the \emph{conciseness} quality attribute. Metrics such as \metric might learn during training that longer methods may need longer summaries and, thus, that very long summaries for short methods may not be appropriate. This is likely aligned with what a developer would think when judging the \emph{conciseness} of a summary. Similarly, \texttt{c\_coeff} is likely to produce low values when a very long summary is associated with a short method, since the denominator of the used formula (\ie the number of terms in the summary) increases. Differently, conventional metrics would judge a generated summary of high quality if it is similar to the reference one. Thus, if the reference is not concise and the generated summary is similar to the reference (thus, not concise as well), the assessed quality will be high and not aligned with the developers' perception.

\begin{figure}[h!]
	\centering
		\caption{Examples from Roy \etal \cite{roy:fse2021} dataset}
	\includegraphics[width=0.98\linewidth]{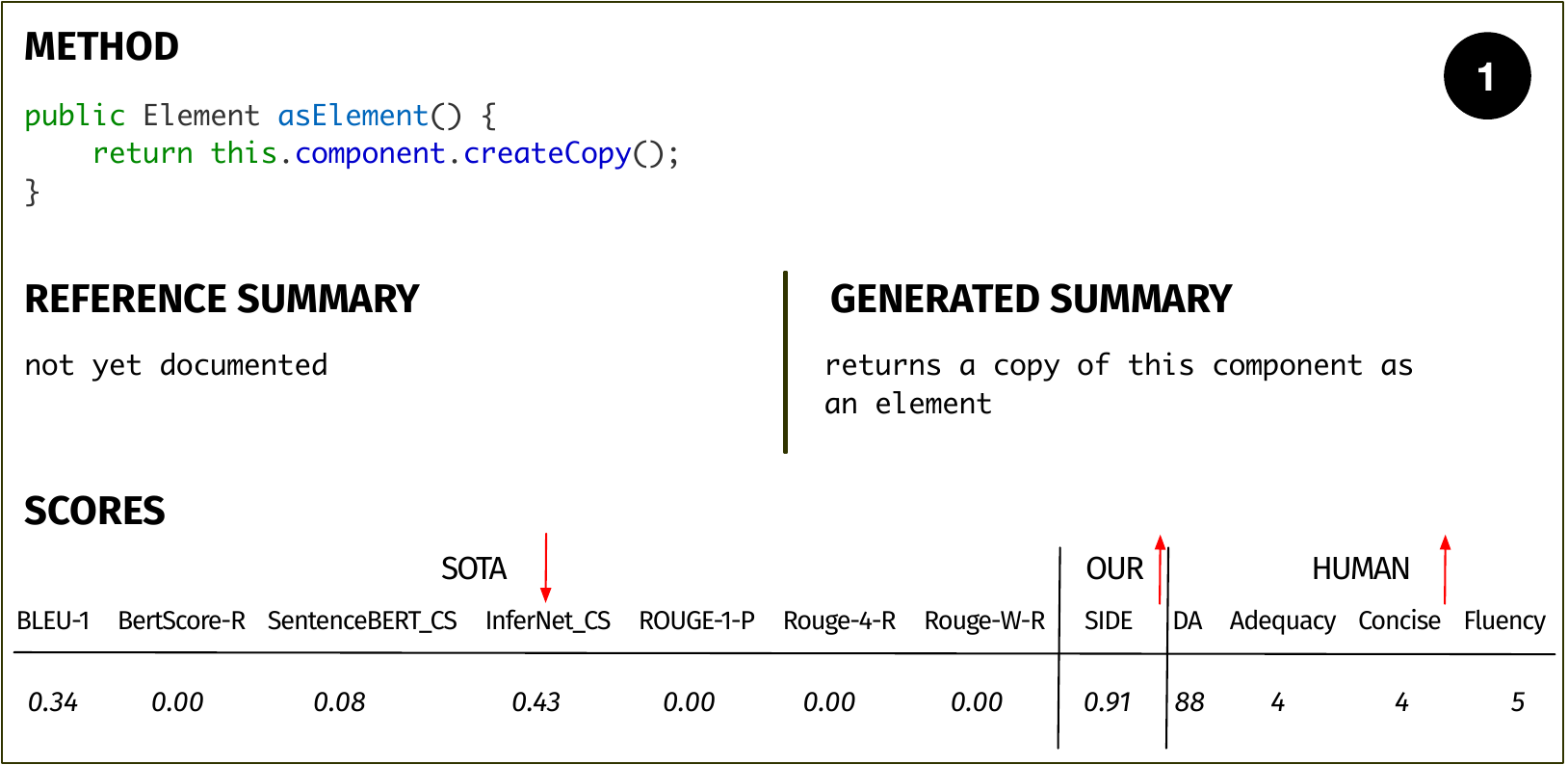}
	\includegraphics[width=0.98\linewidth]{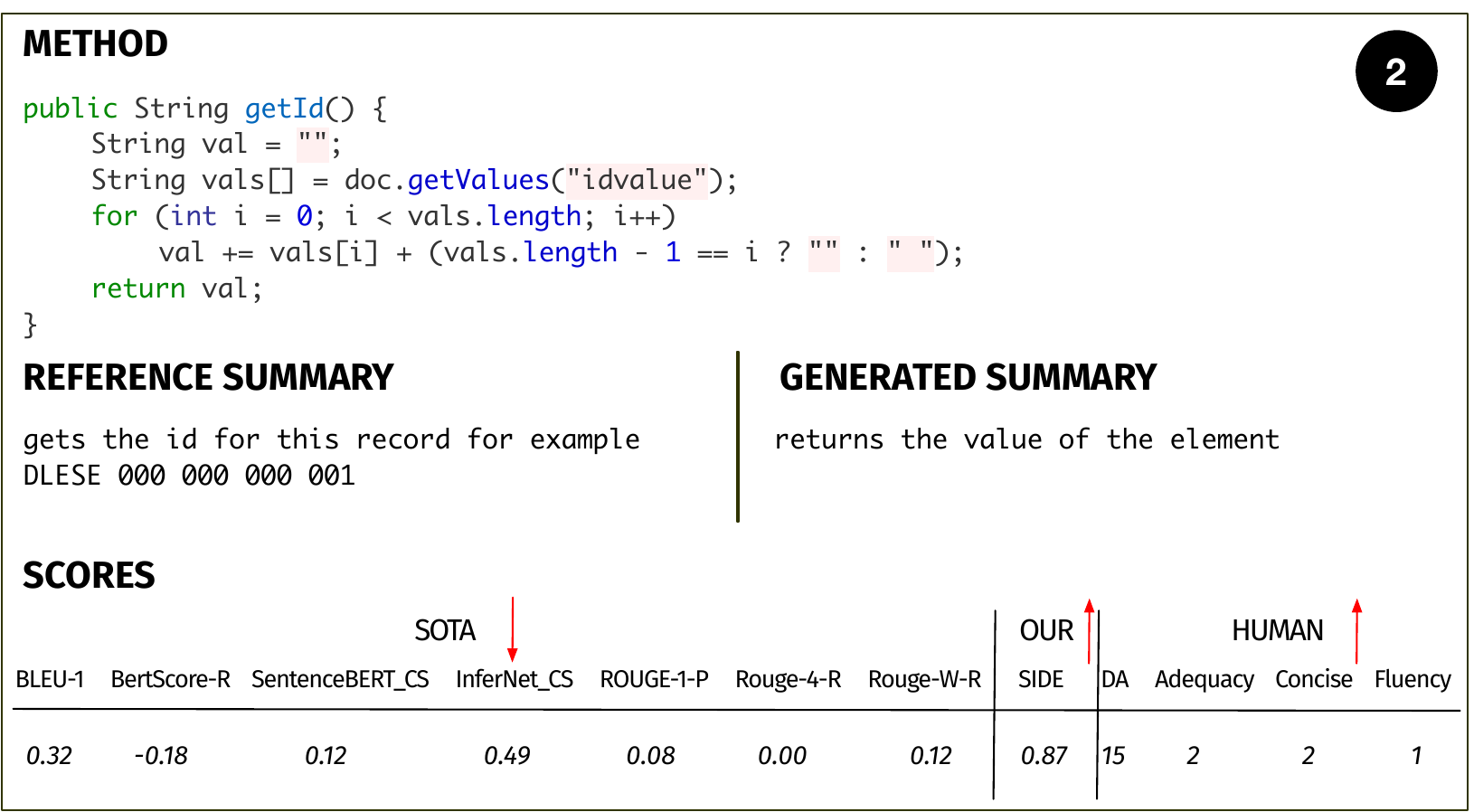}
	\includegraphics[width=0.98\linewidth]{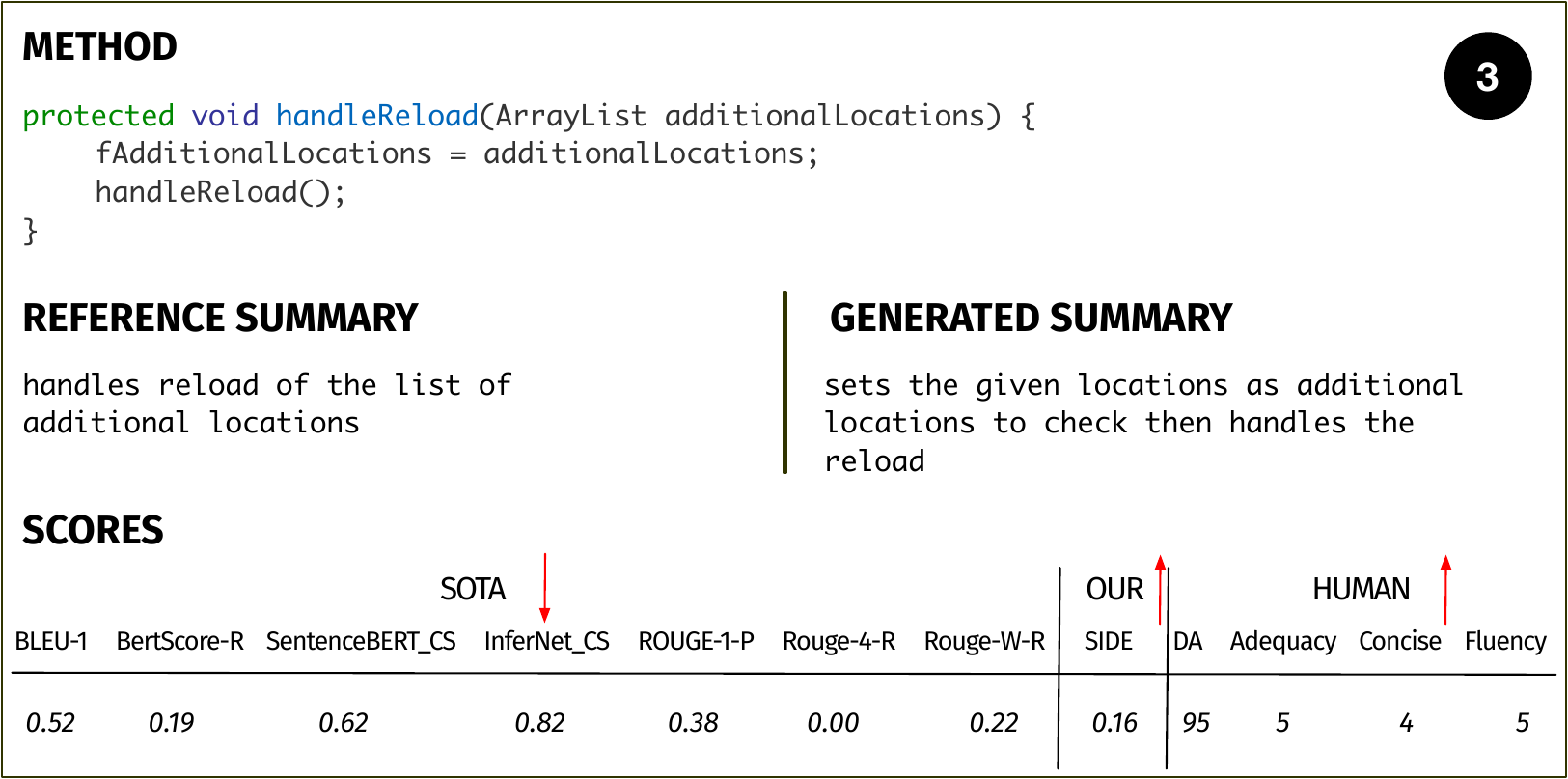}
	\includegraphics[width=0.98\linewidth]{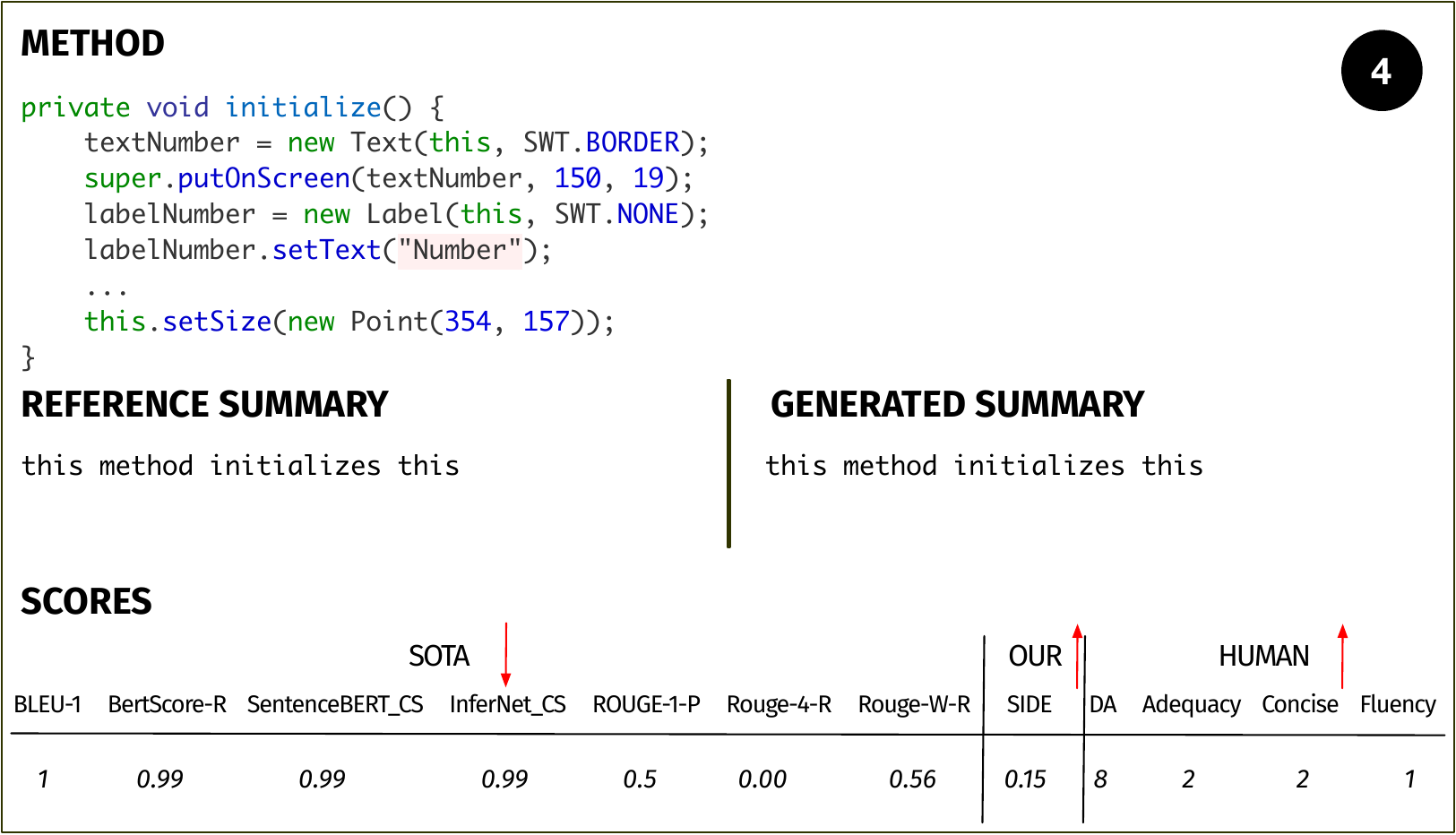}
	\vspace{-0.8cm}
	\label{fig:qualitative}

\end{figure}

\subsection{Qualitative Analysis}

To provide further insights into the complementarity of the two families of metrics, we selected four qualitative examples to discuss (\figref{fig:qualitative}). Each example features (i) the Java method for which a summary was automatically generated; (ii) the target summary (\ie the one written by developers); (iii) the generated summary; and (iv) the quality scores assigned by the seven state-of-the-art (SOTA) metrics selected via the \emph{redun} procedure (\ie \texttt{BLEU-1}, \texttt{BERTScore-R}, \etc), by our \metric metric, and in the human assessment. We selected two examples in which \metric provides an indication aligned with the developers' perception while the SOTA metrics fail, and two examples in which the opposite scenario occurs. 

The first example \circled{1} is a typical scenario in which the reference summary is of low quality since it actually documents the fact that the method is ``\emph{not yet documented}''. 

The generated summary, instead, provides a good description of the method accordingly to the developers and, thus, is substantially different from the reference. This results in low values for the SOTA metrics, while \metric agrees with the human assessment (\emph{DA score} = 0.88), assigning the generated summary 0.91/1.00.

In the second \circled{2} and third \circled{3} examples, \metric is instead in disagreement with developers. In \circled{2}, it assigns a high score (0.87) to a correct but very generic summary, while the conventional metrics are  able to assess the low quality of the summary thanks to the presence of additional material present in the reference but lacking in the generated summary (\ie the example). 

In \circled{3}, \metric assigns a quite low score to the generated summary, despite it being well-judged by humans. SOTA metrics can instead exploit the word overlap between the generated and the reference summary, assessing the summary with higher scores.

Finally, \circled{4} shows a low-quality generated summary (as assessed by developers) which is scored with the maximum scores by state-of-the-art metrics, since being identical to the reference one. \metric is instead able to identify the summary as low quality (0.15) by contrasting it with the documented code.

In summary, our analysis suggests that a more comprehensive evaluation framework must be considered when assessing the quality of automatically generated summaries: Conventional metrics looking at the similarity between the generated and the reference summary are not enough and should be augmented by metrics judging the suitability of a generated summary for the code it documents, similarly to what \metric does. The latter is the metric more related to the human judgment of code summary quality.

\subsection{Ablation Study - Impact of Hard-negatives}
\label{subsec:ablation}

\rev{
	Due to the lack of space, we present in our replication package \cite{replication} a table reporting the results of the ordered logistic regression model for each dependent variable including two variants of \metric: One trained using hard-negatives (\ie the one related to the results discussed as of now) and one trained without hard-negatives. In short, the impact of including or not hard-negatives in the training set is negligible, and the all findings of our study are not changed by such a design choice (\eg no impact on which ORs are statistically significant nor on their magnitude). This result may be due to several reasons. First, due to our definition of hard-negatives (\ie comments documenting at most 25\% of the method's statements), we managed to create hard-negatives for only $\sim$15\% of the methods in our dataset. Thus, it is possible that a higher coverage is needed to observe any influence on the achieved results. Second, our definition of hard-negatives may be suboptimal, opening to better approaches aimed at creating hard-negative samples for contrastive learning applied to software engineering tasks.
}

%% file: tables/pca.tex


\begin{table}[ht]
	\centering
		\caption{RQ$_1$ - PCA for the evaluated metrics}
		\resizebox{\columnwidth}{!}{
		\label{tab:pca}
				\begin{tabular}{rrrrrrrrrrr}
					\toprule
					& {\bf PC1} & {\bf PC2} & {\bf PC3} & {\bf PC4} & {\bf PC5} & {\bf PC6} & {\bf PC7} & {\bf PC8} & {\bf PC9} & {\bf PC10} \\
					\addlinespace[0.08cm]\midrule\addlinespace[0.08cm]

						{\bf Prop. of Variance} & 0.55 & 0.16 & 0.08 & 0.07 & 0.04 & 0.03 & 0.02 & 0.02 & 0.01 & 0.01 \\ 
  						{\bf Cumulative Prop.} & 0.55 & 0.71 & 0.80 & 0.87 & 0.91 & 0.94 & 0.96 & 0.98 & 0.99 & 1.00 \\ 
						\addlinespace[0.08cm]\midrule\addlinespace[0.08cm]
						{\bf BLEU-1} & 0.26 & 0.24 & -0.44 & 0.11 & \cellcolor[HTML]{656565}\color[HTML]{FFFFFF} -0.50 & 0.13 & -0.44 & 0.45 & -0.09 & -0.09 \\ 
					  	{\bf BERTScore-R} & \cellcolor[HTML]{656565}\color[HTML]{FFFFFF}  0.47 & 0.15 & -0.24 & 0.25 & -0.18 & -0.18 & \cellcolor[HTML]{656565}\color[HTML]{FFFFFF} 0.74 & -0.07 & 0.10 & -0.09 \\ 
					  	{\bf SentenceBERT\_CS} & 0.38 & -0.10 & 0.04 & 0.17 & 0.22 & \cellcolor[HTML]{656565}\color[HTML]{FFFFFF} -0.68 & -0.37 & -0.18 & -0.32 &	 -0.18 \\ 
					  	{\bf InferSent\_CS} & 0.23 & -0.01 & -0.01 & 0.06 & 0.09 & -0.21 & -0.17 & 0.10 & 0.54 & \cellcolor[HTML]{656565}\color[HTML]{FFFFFF}0.75 \\  
					  	{\bf ROUGE-1-P} & \cellcolor[HTML]{656565}\color[HTML]{FFFFFF} 0.44 & 0.07 & -0.10 & 0.21 & 0.33 & \cellcolor[HTML]{656565}\color[HTML]{FFFFFF}0.63 & -0.17 & -0.43 & -0.14 & 0.11 \\ 
					  	{\bf ROUGE-4-R} & 0.41 & 0.26 & 0.38 & \cellcolor[HTML]{656565}\color[HTML]{FFFFFF} -0.73 & -0.25 & 0.01 & -0.02 & -0.14 & -0.05 & 0.02 \\ 
					  	{\bf ROUGE-W-R} & 0.30 & -0.04 & 0.36 & 0.06 & 0.44 & 0.15 & 0.05 & \cellcolor[HTML]{656565}\color[HTML]{FFFFFF} 0.64 & 0.19 & -0.32 \\
						{\bf c\_coeff} & 0.13 & -0.47 & \cellcolor[HTML]{656565}\color[HTML]{FFFFFF} 0.53 & 0.39 & \cellcolor[HTML]{656565}\color[HTML]{FFFFFF} -0.54 & 0.13 & -0.07 & -0.10 & 0.06 & -0.01 \\ 
					  	{\bf CodeT5-plus\_CS} & -0.00 & -0.02 & -0.17 & -0.08 & -0.03 & -0.02 & -0.23 & -0.34 & \cellcolor[HTML]{656565}\color[HTML]{FFFFFF} 0.72 &  -0.53 \\ 
					  	{\bf SIDE} & 0.20 & \cellcolor[HTML]{656565}\color[HTML]{FFFFFF} -0.79 & -0.39 & -0.39 & 0.03 & 0.07 & 0.08 & 0.10 & -0.07 & 0.02 \\ 
					  	\bottomrule
				\end{tabular}
		}
		\vspace{-0.45cm}
\end{table}

%% file: tables/logistic-model.tex
\begin{table}[ht]
	\centering
	\caption{RQ$_2$ - Ordered logistic regression model\vspace{-0.3cm}}
	\resizebox{0.85\columnwidth}{!}{
			\begin{tabular}{lrrrrrr}
				\toprule 
				\multicolumn{6}{c}{	\cellcolor[HTML]{656565}\color[HTML]{FFFFFF} Overall DA Score \cite{roy:fse2021}}\\\midrule 
				\bf Metric & \bf OR & \bf Value & \bf Std. Error & \bf $t$-value & \bf $p$-value \\ 
				\hline
				BLEU-1 & 0.9990 & -0.0010 & 0.0017 & -0.5822 & 0.6222 \\ 
				BERTScore-R & 1.0029 & 0.0029 & 0.0023 & 1.2634 & 0.2943 \\ 
				SentenceBERT\_CS & \cellcolor[HTML]{656565}\color[HTML]{FFFFFF}  1.0057 & 0.0057 & 0.0020 & 2.8882 & 0.0100 \\ 
				InferNet\_CS & 0.9980 & -0.0020 & 0.0025 & -0.8064 & 0.5250 \\ 
				 
				ROUGE-1-P & \cellcolor[HTML]{656565}\color[HTML]{FFFFFF}  1.0058 & 0.0058 & 0.0018 & 3.2552 & 0.0033 \\ 
				ROUGE-4-R & 1.0024 & 0.0024 & 0.0011 & 2.1987 & 0.04667 \\ 
				ROUGE-W-R & 0.9996 & -0.0004 & 0.0017 & -0.2485 & 0.8040 \\
				c\_coeff & \cellcolor[HTML]{656565}\color[HTML]{FFFFFF}  1.0143 & 0.0142 & 0.0012 & 11.4549 & $<$0.0001 \\ 
				CodeT5-plus\_CS & \cellcolor[HTML]{656565}\color[HTML]{FFFFFF}  1.0044 & 0.0044 & 0.0017 & 2.5383 & 0.0220 \\ 
				\metric & \cellcolor[HTML]{656565}\color[HTML]{FFFFFF}  1.0205 & 0.0203 & 0.0017 & 11.7029 & $<$0.0001 \\ 
				\hline
			   & &  &  & &  \\
			   \toprule 
			   	\multicolumn{6}{c}{	\cellcolor[HTML]{656565}\color[HTML]{FFFFFF} Content Adequacy \cite{roy:fse2021}}\\\midrule 
				\bf Metric & \bf OR & \bf Value & \bf Std. Error & \bf $t$-value & \bf $p$-value \\ 
			   	\hline
			   	BLEU-1 & 0.9672 & -0.0333 & 0.0347 & -0.9615 & 0.4200 \\ 
			   	BERTScore-R & 1.0525 & 0.0512 & 0.0468 & 1.0927 & 0.3929 \\ 
			   	SentenceBERT\_CS & \cellcolor[HTML]{656565}\color[HTML]{FFFFFF}  1.1121 & 0.1063 & 0.0406 & 2.6172 & 0.0225 \\ 
			   	InferNet\_CS & 1.0000 & 0.0000 & 0.0517 & -0.0004 & 1.0000 \\ 
			   	 
			   	ROUGE-1-P & \cellcolor[HTML]{656565}\color[HTML]{FFFFFF}  1.0943 & 0.0901 & 0.0364 & 2.4733 & 0.0260 \\ 
			   	ROUGE-4-R & 0.9914 & -0.0086 & 0.0226 & -0.3814 & 0.7811 \\ 
			   	ROUGE-W-R & 1.0437 & 0.0428 & 0.0351 & 1.2202 & 0.3700 \\ 
				c\_coeff & \cellcolor[HTML]{656565}\color[HTML]{FFFFFF}  1.3729 & 0.3169 & 0.0255 & 12.4350 & $<$0.0001 \\
			   	CodeT5-plus\_CS & \cellcolor[HTML]{656565}\color[HTML]{FFFFFF}  1.1413 & 0.1322 & 0.0358 & 3.6877 & $<$0.0001 \\ 
			   	\metric & \cellcolor[HTML]{656565}\color[HTML]{FFFFFF}  1.6265 & 0.4864 & 0.0366 & 13.2942 & $<$0.0001 \\ 
			   	\hline
			   	 & &  &  & &  \\
			   	\toprule 
			   	\multicolumn{6}{c}{	\cellcolor[HTML]{656565}\color[HTML]{FFFFFF} Conciseness \cite{roy:fse2021}}\\\midrule 
				\bf Metric & \bf OR & \bf Value & \bf Std. Error & \bf $t$-value & \bf $p$-value \\ 
				\hline
				BLEU-1 & 1.0656 & 0.0636 & 0.0342 & 1.8576 & 0.1575 \\ 
				BERTScore-R & 1.0160 & 0.0159 & 0.0470 & 0.3375 & 0.8770 \\ 
				SentenceBERT\_CS & 1.0587 & 0.0571 & 0.0403 & 1.4163 & 0.2617 \\ 
				InferNet\_CS & 1.0079 & 0.0079 & 0.0511 & 0.1547 & 0.8770 \\ 
				ROUGE-1-P & 1.0075 & 0.0075 & 0.0359 & 0.2091 & 0.8770 \\ 
				ROUGE-4-R & 0.9937 & -0.0063 & 0.0228 & -0.2766 & 0.8770 \\ 
				ROUGE-W-R & \cellcolor[HTML]{656565}\color[HTML]{FFFFFF}  1.0954 & 0.0911 & 0.0346 & 2.6352 & 0.0267 \\ 
				c\_coeff & \cellcolor[HTML]{656565}\color[HTML]{FFFFFF}  1.2433 & 0.2178 & 0.0254 & 8.5798 & $<$0.0001 \\ 
				CodeT5-plus\_CS & 1.0630 & 0.0611 & 0.0355 & 1.7212 & 0.1700 \\ 
				\metric & \cellcolor[HTML]{656565}\color[HTML]{FFFFFF} 1.3844 & 0.3253 & 0.0354 & 9.1835 & $<$0.0001 \\ 
				\hline
				 & &  &  & &  \\
				\toprule 
				\multicolumn{6}{c}{	\cellcolor[HTML]{656565}\color[HTML]{FFFFFF} Fluency \cite{roy:fse2021}}\\\midrule 
				\bf Metric & \bf OR & \bf Value & \bf Std. Error & \bf $t$-value & \bf $p$-value \\ 
				\hline
				BLEU-1 & 1.0650 & 0.0629 & 0.0345 & 1.8243 & 0.2267 \\ 
				BERTScore-R & 1.0559 & 0.0544 & 0.0467 & 1.1638 & 0.4600 \\ 
				SentenceBERT\_CS& 0.9958 & -0.0042 & 0.0404 & -0.1043 & 0.9170 \\ 
				InferNet\_CS & 1.0543 & 0.0529 & 0.0515 & 1.0272 & 0.4600 \\ 
				ROUGE-1-P & 1.0334 & 0.0329 & 0.0365 & 0.9004 & 0.4600 \\ 
				ROUGE-4-R & 1.0304 & 0.0299 & 0.0229 & 1.3061 & 0.4600 \\ 
				ROUGE-W-R & 0.9672 & -0.0334 & 0.0347 & -0.9603 & 0.4600 \\ 
				c\_coeff & \cellcolor[HTML]{656565}\color[HTML]{FFFFFF}  1.1680 & 0.1553 & 0.0253 & 6.1437 & $<$0.0001 \\ 
				CodeT5-plus\_CS & 1.0249 & 0.0246 & 0.0357 & 0.6902 & 0.5444 \\ 
				\metric & \cellcolor[HTML]{656565}\color[HTML]{FFFFFF} 1.2826 & 0.2489 & 0.0359 & 6.9359 & $<$0.0001 \\ 
				\bottomrule
			\end{tabular}
		}
		\label{tab:logistic}
		\vspace{-0.3cm}
\end{table}

%% file: threats.tex
\section{Threats to Validity} \label{sec:threats}

\emph{Construct validity threats}.  As also detailed in the work of Roy \etal~\cite{roy:fse2021}, the considered dependent variables reflect the user's assessment of a summary from different perspectives (\emph{conciseness}, \emph{fluency}, \emph{content adequacy}, plus an \emph{overall assessment}). We are aware that such an assessment could suffer from the assessor's error or subjectiveness. 
\rev{Also, we acknowledge that the fine-tuning dataset used to train \metric may feature some low-quality summaries which may partially hinder its ability to assess whether a natural language text represents a suitable summary for a given code. However, our empirical evaluation showed that \metric is still the metric better capturing human assessment of code summary quality, despite this potential source of noise in its training set.}

\emph{Internal validity threats}. These primarily pertain to the configurations applied during metric computation and evaluation, such as the decision to utilize cosine similarity. In addition, we recognize the potential value in exploring alternative hyperparameter settings (see \secref{sub:training-evaluation}) when developing \metric. 

Also, adopting CodeT5+ as the representative state-of-the-art pre-trained model for generating $\langle method, documentation \rangle$ embeddings, and subsequently using cosine similarity for comparison, might not reflect the best strategy for the discussed task.

\rev{Finally, the negligible impact of including hard-negative samples in the training set may be due to the design we employed to mine hard-negatives (\eg see the choice of the 25\% coverage to identify hard-negative comments) and further studies are needed to investigate alternative solutions.
}

\emph{Conclusion validity threats}. RQ$_1$ findings show both the metrics correlation, computed using a non-parametric procedure, and the PCA. RQ$_2$ is based on an ordered logistic regression, suitable for variables in ordinal scale. Note that statistically significant $p-$values are small enough to be unaffected by fishing and error rate. 

\emph{External validity threats}. The main limitation of our study is that it relies on human evaluations from a single dataset \cite{roy:fse2021}. Nevertheless, such a dataset is large and features a total of 6,253 (5,201 used in our study) evaluations from 226 different participants.
Concerning the independent variables (evaluation metrics), our study attempted to consider several metrics having a different nature, some of which were used in previous work \cite{haque:icpc2022,roy:fse2021,Steidl:icpc2013}. 

\vspace{-0.20cm}

%% file: conclusion.tex
\section{Conclusions} \label{sec:conclusion}

State-of-the-art metrics used to evaluate code summarization techniques only assess the similarity between the generated summary and the reference one written by developers. This implies that a generated summary, while suitable for the code to document, may be different from the reference one (\eg a low-quality one), and therefore scored low by these metrics. We argue that the code to document must also be considered in the equation when scoring automatically generated summaries, to assess whether they are suitable for such a code independently from their similarity with the reference summary. To capture this information, we presented \metric, a metric exploiting contrastive learning to learn characteristics of suitable and unsuitable summaries for a given code. 

We run a study involving 40 metrics (including \metric) investigating their complementarity and the extent to which they correlate with humans' evaluation of summary quality. Our findings highlight the high complementarity between \metric and the metrics focusing on the similarity between the generated and the reference summary. \metric is also the metric that better describes humans' assessment of summary quality. Also, we show that the proposed contrastive learning-based metric captures the suitability of a summary for a given code better than simpler solutions.

\textcolor{black}{Further evaluations of code summarization methods should incorporate a broader assessment framework that includes both conventional metrics---\eg the similarity between generated and reference summaries---and \metric-like metrics, evaluating the suitability of the generated summary for the code being documented. In addition, the findings of this study provide several opportunities for researchers, particularly in the area of program comprehension. To this end, a key application of \metric would involve the identification of inconsistencies between code comments with respect to the documented code (\eg a method), as this could significantly improve the quality of software systems once the code and its documentation are realigned.}

%% file: availability.tex
\section{Data Availability} \label{sec:data-availability}

We provide \cite{replication}: (i) the datasets used to train and evaluate \metric; (ii) the code for reproducing the experiments; (iii) the code for computing the \emph{evaluation} metrics; (iv) \texttt{R} scripts for conducting the statistical analysis;  (v) additional results not featured in the paper (\eg correlation matrices); and (vi) a script allowing other researchers to compute \metric for pairs of $\langle$$method$, $summary$$\rangle$.